\documentclass[aps, prl, twocolumn, superscriptaddress, 10pt, nofootinbib]{revtex4-1}
\usepackage[sort&compress]{natbib}   
\usepackage{amsmath,amssymb,amsfonts}       
\usepackage{graphicx}                       

\usepackage[dvipsnames]{xcolor}                        
\usepackage[colorlinks=true]{hyperref}    
\usepackage[UKenglish]{babel}
\usepackage[UKenglish]{isodate}
\usepackage{multirow}
\usepackage{braket}

\usepackage{enumitem}

\newcommand{\gtwo}{\mathit{g}^{(2)}}

\begin{document}
	\title{Hong-Ou-Mandel Interference with Imperfect Single Photon Sources}

\author{H. Ollivier$^\P$}
\author{S. E. Thomas$^\P$}
\email{sarah.thomas@c2n.upsaclay.fr}
\affiliation{Centre for Nanosciences and Nanotechnology, CNRS, Universit\'e Paris-Saclay, UMR 9001,
10 Boulevard Thomas Gobert, 91120, Palaiseau, France}
\author{S. C. Wein}
\affiliation{Institute for Quantum Science and Technology and Department of Physics and Astronomy,
University of Calgary, Calgary, Alberta, Canada T2N 1N4}
\author{I. Maillette de Buy Wenniger }
\author{N. Coste}
\author{J. C. Loredo}
\affiliation{Centre for Nanosciences and Nanotechnology, CNRS, Universit\'e Paris-Saclay, UMR 9001,
10 Boulevard Thomas Gobert, 91120, Palaiseau, France}
\author{N. Somaschi}
\affiliation{Quandela SAS, 10 Boulevard Thomas Gobert, 91120, Palaiseau, France}
\author{A. Harouri}
\author{A. Lemaitre}
\author{I. Sagnes}
\affiliation{Centre for Nanosciences and Nanotechnology, CNRS, Universit\'e Paris-Saclay, UMR 9001,
10 Boulevard Thomas Gobert, 91120, Palaiseau, France}
\author{L. Lanco}
\affiliation{Centre for Nanosciences and Nanotechnology, CNRS, Universit\'e Paris-Saclay, UMR 9001,
10 Boulevard Thomas Gobert, 91120, Palaiseau, France}
\affiliation{Universit\'e Paris Diderot - Paris 7, 75205 Paris CEDEX 13, France}
\author{C. Simon}
\affiliation{Institute for Quantum Science and Technology and Department of Physics and Astronomy,
University of Calgary, Calgary, Alberta, Canada T2N 1N4\\
$^\P$ These authors contributed equally to this work}
\author{C. Anton}
\author{O. Krebs}
\author{P. Senellart}
\email{pascale.senellart-mardon@c2n.upsaclay.fr}
\affiliation{Centre for Nanosciences and Nanotechnology, CNRS, Universit\'e Paris-Saclay, UMR 9001,
10 Boulevard Thomas Gobert, 91120, Palaiseau, France}

	\date{\today}
	
	\begin{abstract}

Hong-Ou-Mandel interference is a cornerstone of optical quantum technologies.  We explore both theoretically and experimentally how the nature of unwanted multi-photon components of single photon sources affect the interference visibility. We apply our approach to quantum dot single photon sources in order to access the mean wavepacket overlap of the single-photon component - an important metric to understand the limitations of current sources. We find that the impact of multi-photon events has thus far been underestimated, and that the effect of pure dephasing is even milder than previously expected.

	\end{abstract}
	
	\maketitle

Quantum interference of indistinguishable single photons is a critical element of quantum technologies. It allows the implementation of logical photon-photon gates for quantum computing~\cite{KLM2001,OBrien2007} as well as the development of quantum repeaters for secure long distance communications~\cite{DLCZ2001,Sangouard2011}. The development of efficient sources of single and indistinguishable photons has become a challenge of the utmost importance in this regard, with two predominant, distinct approaches. The first one is based on non-linear optical photon pair production~\cite{Kwiat1999,Zhong2018}, and multiplexing of heralded single photon sources is being explored to overcome an intrinsic inefficiency~\cite{Kaneda2019,Joshi2018,Xiong2016,Francis-Jones:16}. The other is based on single quantum emitters such as semiconductor quantum dots~\cite{Aharonovich2016,Senellart2017} where ever-growing control of the solid-state emitter has enabled the combination of high efficiency and high indistinguishability~\cite{Somaschi2016,Wang2016,He2019,Wang2019}.  

The standard method to quantify the indistinguishability of single-photon wavepackets is to perform Hong-Ou-Mandel (HOM) interference~\cite{HOM}. In perfect HOM interference, two indistinguishable single photons incident at each input of a 50:50 beam-splitter will exit the beam-splitter together, resulting in no two-photon coincidental detection events at both outputs.  In practice, however, the two inputs only exhibit partial indistinguishability described by a non-unity mean wave-packet overlap $M_\mathrm{s}$ (also defined as the single-photon trace purity~\cite{Fischer2018,Trivedi2020}). Partial indistinguishability of the input states leads to coincidental detection events at the outputs and reduces the HOM interference visibility. The interference visibility $V_\mathrm{HOM}$ can therefore give direct access to the single-photon indistinguishability $M_\mathrm{s}=V_\mathrm{HOM}$ \cite{Trivedi2020}. 

For non-ideal single-photon sources, for which the photonic wavepackets present a residual multi-photon component, the HOM visibility remains the relevant quantity that determines the quality of the above-mentioned quantum operations. However, the visibility of HOM interference is reduced due to multi-photon contributions, even if $M_\mathrm{s}=1$, i.e. for an ideal single-photon indistinguishability. In most cases, the multi-photon component of the photonic wavepacket, characterized by the second order intensity autocorrelation at zero time delay $\gtwo(0)$,  depends on the system parameters in a manner that is completely independent of the single photon indistinguishability, and it is critical to have tools to access the latter in order to understand the physics at play and improve the performance of single photon sources.

Here we explore both theoretically and experimentally HOM interference with imperfect single-photon sources. Previously, the impact of multi-photon contributions on HOM interference has been investigated in the limited case where the additional photons are in the same spectral and temporal mode as the predominant ones~\cite{Bennett2009,Polyakov2011,Huber2017}. It has been shown that the visibility of HOM interference in this case is given by  $V_\mathrm{HOM}  = M_\mathrm{tot} -  \gtwo(0)$~\cite{Uren2005,Trivedi2020}, where $M_\mathrm{tot}$ is the mean wavepacket overlap of the total input state, i.e. including the multi-photon component.  Here we show that the properties of the additional or ``noise" photons play a critical role in HOM interference, and that it is crucial to know the origin of the imperfections to be able to correctly extract the intrinsic single-photon indistinguishability $M_\mathrm{s}$. We validate our approach by experimentally emulating two types of imperfect sources. Finally, we investigate the case of quantum-dot based single photon sources (QDSPS) based on both neutral and charged excitons. By understanding the physical mechanisms in both cases, we are able to provide a proper way of extracting the single photon indistinguishability that accounts for the nature of the multi-photon events.

	\begin{figure*}[t]
	\includegraphics[width=\linewidth]{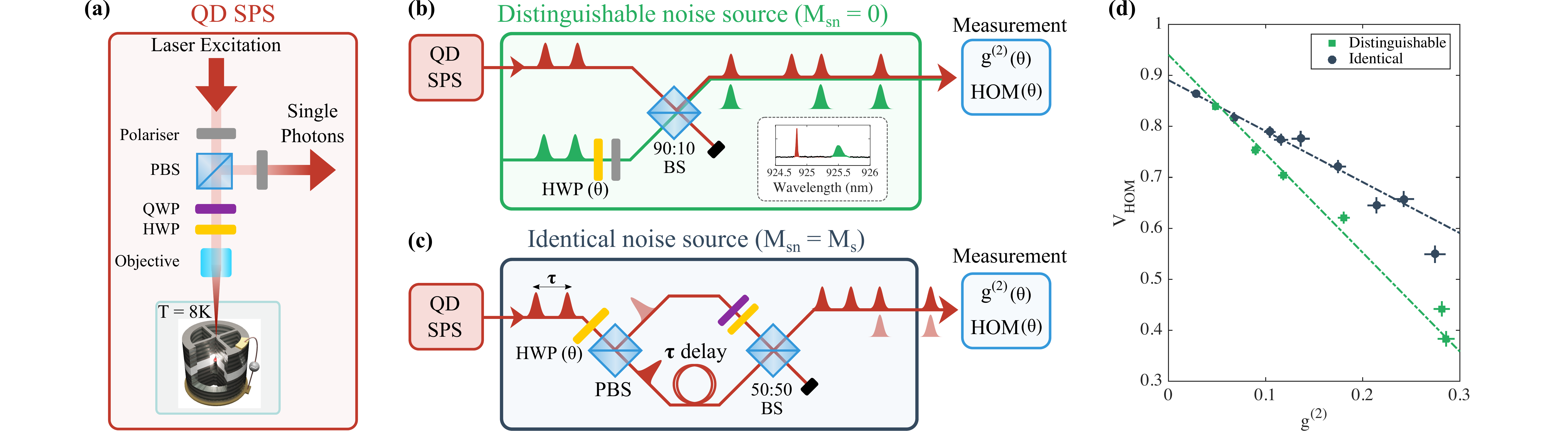}
	\caption{ 	{{ (a) Schematic for the experimental set-up of a quantum dot single photon source (QDSPS). A quantum dot embedded in a micropillar cavity is placed in a cryostat and cooled to around 8K. A resonant excitation pulse is used to coherently control the quantum dot, and single photons are collected in the orthogonal polarization to extinguish the excitation laser.  The half and quarter waveplates allow alignment of the polarization along one of the microcavity axes~\cite{Somaschi2016,Ollivier2020}. (b),(c) Experimental set-up used to emulate an imperfect single photon source with (b) distinguishable ($M_\mathrm{sn} = 0$) and (c) identical noise $M_\mathrm{sn} = M_\mathrm{s}$.  The inset in (b) shows the spectrum of the spectrally-distinct noise photons and QD photons. (d) Visibility of HOM interference, $V_\mathrm{HOM}$ measured as a function of $\gtwo$ for distinguishable (green squares) and identical (black circles) noise sources. The lines are the predictions from the theoretical model.}} \label{Fig_LimitingCases}}
\end{figure*}

We model an imperfect ``single-photon"  state ($\gtwo(0) > 0$) by mixing a true single photon ($\gtwo(0) =0$) with separable noise at a beam splitter. We limit our analysis to small $\gtwo(0)$ values so that the noise field itself is well-approximated by an optical field with at most one additional photon and a large vacuum contribution. This restriction to a weak, separable noise field remains relevant in practice for many situations as illustrated for QDSPSs later on.

It can be shown (see Supplementary Material), that for separable noise and a small resultant $\gtwo$ (typically $\gtwo <0.3$), the visibility of HOM interference is given by: 
\begin{equation}
    V_\mathrm{HOM} = M_\mathrm{s} - \left( \frac{1 + M_\mathrm{s}}{1 + M_\mathrm{sn}} \right) g^{(2)} \label{Eqn_Model}
\end{equation}

\noindent where $M_\mathrm{sn}$ is the mean wavepacket overlap between the single photon and an additional noise photon satisfying $0 \le M_\mathrm{sn} \le M_\mathrm{s}$. We have defined  $\gtwo \equiv \gtwo(0)$ for simplicity.

It is instructive to consider the two limiting cases of Equation~\ref{Eqn_Model}. If the additional photons are identical to the single photons, i.e. $M_\mathrm{sn} = M_\mathrm{s}$, then Equation~\ref{Eqn_Model} reduces to the simple case that $ V_\mathrm{HOM} = M_\mathrm{s} - \gtwo$, showing that the total and single photon mean wavepacket overlap coincide, $M_\mathrm{s}=M_\mathrm{tot}$. Alternatively, if the noise has no overlap with the single photons and $M_\mathrm{sn} = 0$ , then the visibility is further reduced and given by $V_\mathrm{HOM} = M_\mathrm{s} - \left( {1 + M_\mathrm{s}} \right) g^{(2)} $. The degree to which HOM interference is affected by a non-zero $\gtwo$ is therefore dependent on the origin of the additional photons.

We experimentally test this model by emulating imperfect single photon sources. We prepare a train of near-optimal single photons and mix them with additional photons to controllably increase $\gtwo$ and measure the impact on the HOM interference. We experimentally emulate the two limiting cases outlined above: when the additional photons are completely \textit{distinguishable} ($M_\mathrm{sn} = 0$) from our single photon input, and when they are completely \textit{identical} ($M_\mathrm{sn} = M_\mathrm{s}$). In each case, we measure the $\gtwo$ and HOM interference visibility of the resultant wavepacket in an unbalanced Mach-Zehnder interferometer (see Supplementary Material for further details).

We use a state-of-the-art single photon source based on a quantum dot (QD) deterministically embedded in an electrically contacted micropillar cavity~\cite{Somaschi2016}.  The QD acts as an artificial atom which we coherently control via resonant excitation to generate single photons with high single photon purity, $\gtwo<0.05$, and high indistinguishability, $M_\mathrm{tot}>0.9$. The single photons are separated from the excitation laser using a cross-polarization set-up, as shown in Figure~\ref{Fig_LimitingCases}(a).

The experimental set-ups that enable a controlled increase of the multi-photon probability are shown in Figure~\ref{Fig_LimitingCases}(b) and (c) for the two limiting cases. First, to add fully distinguishable photons, we mix the single photons from the QDSPS with attenuated laser pulses at a different wavelength. A 3~ps Ti-Sapph pulsed laser centered at 925 nm is spectrally dispersed using a diffraction grating, and a narrow portion is selected to obtain a 15~ps  excitation pulse resonant with the QD transition (here a charged exciton). A second, non-overlapping part of the spectrum is selected to mix with the emitted single photons. By appropriately tuning the time delay we can add synchronous spectrally-distinguishable photons to the single photon emission. The corresponding output field is then considered as an effective source, and we adjust the power of the laser beam to alter the magnitude of the two-photon component. The measured HOM visibility as a function of $\gtwo$ of this effective source is shown in Figure~\ref{Fig_LimitingCases}(d). Since the spectral overlap between the QDSPS photons and the additional laser photons is zero ($M_{\mathrm{sn}}= 0$) our model predicts that $V_\mathrm{HOM} =  M_\mathrm{s} - (1 +  M_\mathrm{s})\gtwo$,  where a single parameter $M_\mathrm{s}$ accounts for both the origin of the curve at zero $\gtwo$ and its slope. The line in Figure~\ref{Fig_LimitingCases}(d) shows that this model fits the data very well with $M_\mathrm{s}=0.94 \pm 0.02 $.

To create a wavepacket where the additional photons are identical to the predominant single photon component, we build another effective source where we add a small fraction of photons from the same QDSPS generated at a later time. This is obtained by performing an unbalanced quantum interference between two photon pulses produced by the QDSPS with delay $\tau$.  The first half waveplate (HWP) and polarizing beam splitter (PBS) in Figure~\ref{Fig_LimitingCases}(c) allow us to tune the relative intensity of the predominant single photon pulse and the additional photons. Then, a pair of QWP and HWP is used to make the polarizations of both photons identical before the 50:50 beam splitter (BS). Most of the time, only the main photon gets to the second BS and is transmitted with 50\% probability. However, when this photon meets a second one generated after a $\tau$ delay and temporally overlapped, they will undergo HOM interference and  exit the beam splitter in the same output port. Therefore, the output of the second BS has a higher $\gtwo $, since there is a small probability that some of the output pulses now contain two identical photons. By adjusting the splitting ratio at the first beam splitter, the  $\gtwo$ of the output state can be controlled. Figure~\ref{Fig_LimitingCases}(d) presents the HOM visibility as we increase the $\gtwo$ via addition of identical photons, where a clear difference is observed compared to the previous limiting case. For $ M_\mathrm{sn} = M_\mathrm{s}$, the model predicts $V_\mathrm{HOM} =  M_\mathrm{s} - \gtwo$, a linear dependence with slope of $-1$. The line in Figure~\ref{Fig_LimitingCases}(f) again demonstrates that the model gives a very good fit to the data, with an extracted $M_\mathrm{s} = 0.89 \pm 0.01 $. 

We note that the extracted values of $M_\mathrm{s}$ for these two cases represent the upper and lower bound of the intrinsic single photon indistinguishability of the QDSPS used in these measurements. If the non-zero $\gtwo$ of the QDSPS was due to distinguishable noise then we could deduce that $M_\mathrm{s}=0.94 \pm 0.02 $. Similarly if the noise was identical then the QDSPS has a single photon indistinguishability of $M_\mathrm{s} = 0.89 \pm 0.01 $. This demonstrates that it is necessary to know the origin of the unwanted photon emission in order to be able to extrapolate the data back to $\gtwo = 0$.

Our study highlights the importance of determining the origin of the multi-photon component in order to properly extract the single photon indistinguishability. To the extent of our knowledge, this has so far only been done in the indistinguishable case, independent of the physical phenomena of the multi-photon components. In the following, we discuss how to properly estimate the single photon indistinguishability for the current highest performing single photon sources, i.e. QD based sources.

There are two distinct categories of QDSPS, depending on the charge state of the quantum dot: neutral excitons and charged excitons (hereafter referred to as exciton and trion states respectively). The optical selection rules and photon emission processes differ significantly between the excitons and trions~\cite{Ollivier2020}, leading to a different origin of the multi-photon component.

	\begin{figure}
	\includegraphics[width=\linewidth]{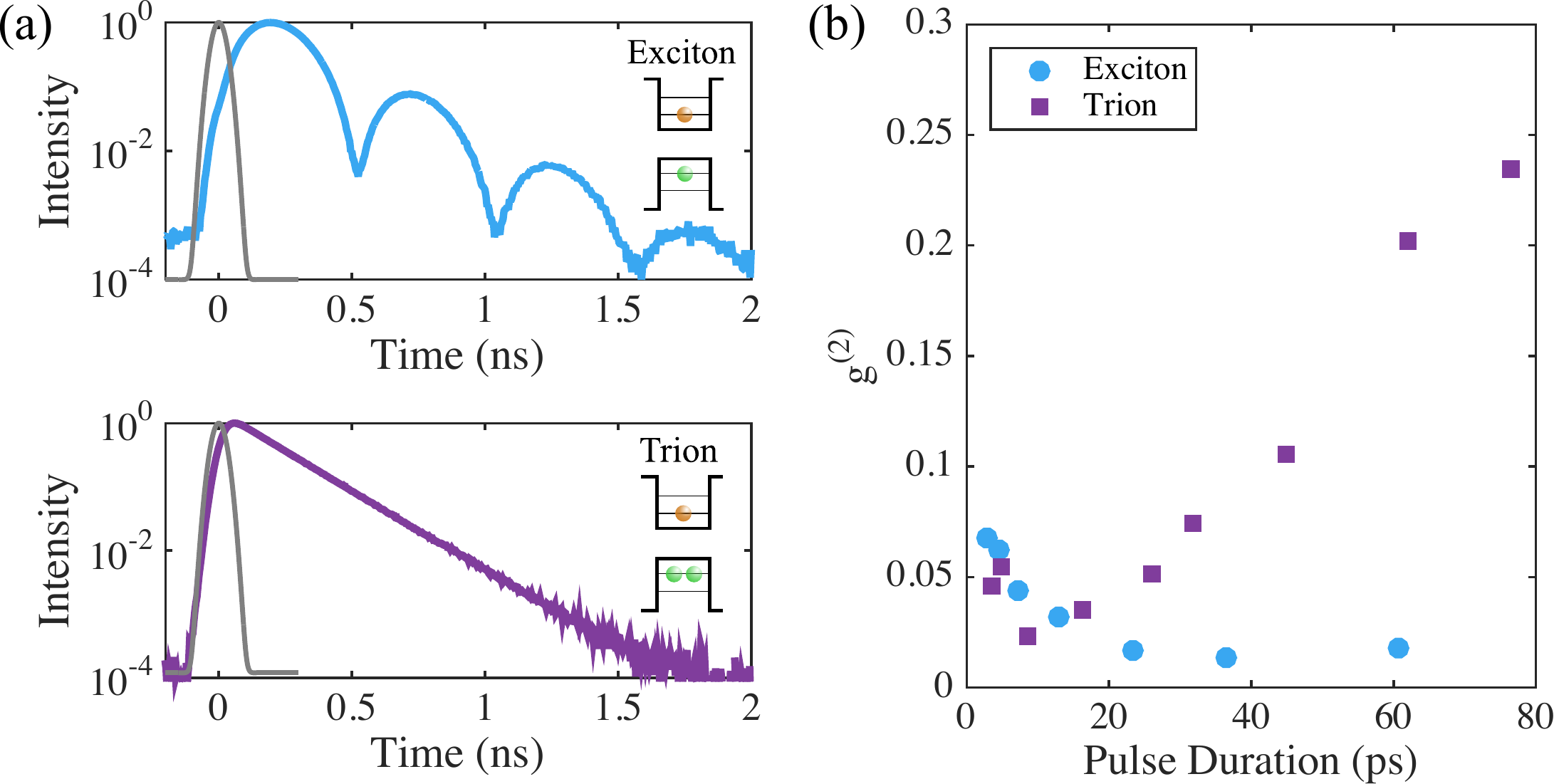}
	\caption{ 	{(a) Time traces of the single photon wavepacket emitted by exciton (upper) and trion (lower) based sources. The excitation laser pulse is shown in grey. (b) Measured $\gtwo$ for trion (purple squares) and exciton (blue circles) based sources as a function of the excitation pulse duration {at $\pi-$pulse}. The error bars are within the size of the plotted points.  } \label{Fig_Exciton_Trion_Lifetime}}
\end{figure}

For an exciton, the system is described by a three level system where the excitation pulse creates a superposition of the two excitonic linear dipoles with an energy difference given by the fine-structure splitting~\cite{Bayer2002}. This results in a time dependent phase between the two exciton eigenstates, so that the single photon emission in cross-polarization beats with a period determined by the fine structure splitting~\cite{Lenihan2002,Ollivier2020}, as shown in Figure~\ref{Fig_Exciton_Trion_Lifetime}(a). These optical selection rules imply that the single photon emission in cross polarisation is delayed with respect to the excitation pulse. For a trion based source, the optical selection rules correspond to a four level-system with four possible linearly polarized transitions~\cite{Xu2007}. In a cross-polarised set-up, this system behaves like an effective two-level system, and the single photon emission shows a rapid rise time and monoexponential decay as shown in Figure~\ref{Fig_Exciton_Trion_Lifetime}(a).

These optical selection rules result in very different origins of the residual two-photon component for the two types of sources. To illustrate this, we measure the $\gtwo$ at maximum emitted brightness for two QD sources, one exciton and one trion, whilst increasing the temporal length of the excitation pulse (Figure~\ref{Fig_Exciton_Trion_Lifetime}(b)). For the exciton source,  the single photon purity remains very high for pulse durations up to 80~ps, whereas the single photon purity rapidly degrades for longer pulses for the trion based source. For the trion, the single photon emission process is fast and can occur during the laser pulse so that there is a probability that the quantum dot returns to the ground state before the end of the laser pulse and gets excited again, leading to the emission of a second photon ~\cite{Fischer2017}. For the exciton, the delayed emission in cross polarization means that the probability of collecting two photons via re-excitation is very small, and the measured value of $\gtwo$ for the exciton remains small for pulse durations of up to 80~ps. We notice that for both excitons and trions, the $\gtwo$ is higher for very short pulses.  This is because the power required to reach maximum emitted brightness ($\pi$-pulse) increases as the pulse duration decreases~\cite{Giesz2016}. This implies that in the very short pulse regime ($ < 10$~ps) , the $\gtwo$ is limited by imperfect suppression of the excitation laser. This remains the dominant source of an imperfect $\gtwo$ for exciton sources up to a pulse duration of 80~ps, whereas trion sources are limited by re-excitation for pulses longer than 15~ps. 

To correctly extract the single photon indistinguishability for each type of QDSPS it is critical to account for these different origins of the multi-photon component. To do so, we experimentally increase the multi-photon component by adjusting the main parameter that is responsible for multi-photon emission in each type of source, and then measure the impact this has on the visibility of HOM interference. Specifically, we increase the probability of re-excitation for a trion source and the amount of laser photons for the exciton source.

\begin{figure}
	\includegraphics[width=\linewidth]{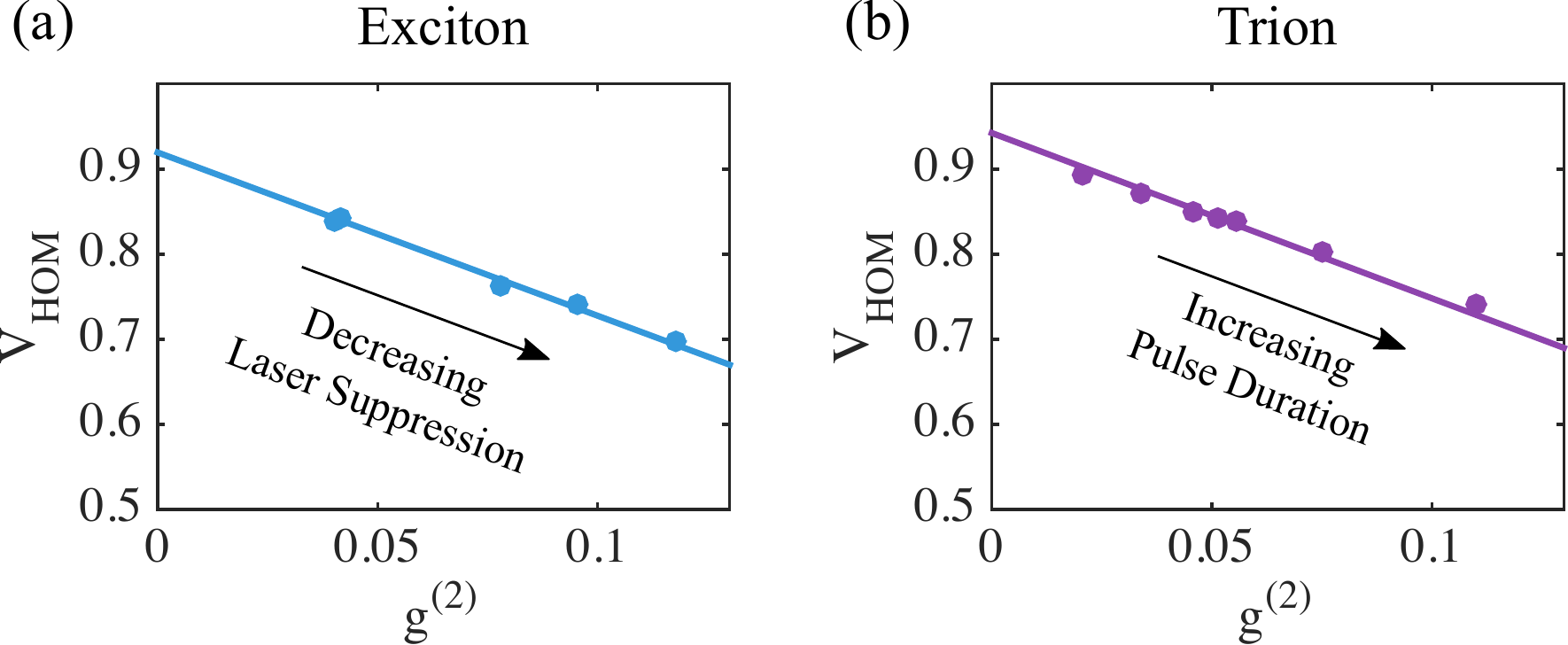}
	\caption{ 	{ (a)  Measured HOM visibility as a function of the $\gtwo$ for a exciton source, as the suppression of the excitation laser is worsened to increase the $\gtwo$.  (b) Measured HOM visibility as a function of the $\gtwo$ for a trion source as the pulse duration is increased. For longer pulse durations, there is more re-excitation and the $\gtwo$ is higher. Each data point was taken at ``$\pi$-pulse", corresponding to  maximum brightness of the source. In both plots, the solid line gives the theoretical prediction for these data. The error bars are within the size of the plotted points.} \label{Fig_TrionExciton}}

\end{figure}

For the exciton based source, we add laser photons to the single photon emission from the QDSPS by turning the quarter waveplate (QWP) of the excitation pulse (see Figure~\ref{Fig_LimitingCases}(a)). This means that the excitation pulse is no longer aligned along one of the polarisation axes of the cavity, and the light will experience a small amount of polarisation rotation due to the birefringence of the cavity. Therefore, some fraction of the excitation pulse will now be collected in the orthogonal polarisation with the single photons. By adjusting the QWP we can add more laser photons and increase the $\gtwo$ of this effective source and measure the corresponding impact on HOM interference, as shown in Figure~\ref{Fig_TrionExciton}(a). The added noise photons from the laser are separable from the single photons and therefore Equation~\ref{Eqn_Model} can be used to model the data.  From the time traces in Figure~\ref{Fig_Exciton_Trion_Lifetime}(b), we can calculate that there is very little overlap between the laser photons and the single photons emitted by the quantum dot so that $M_\mathrm{sn} \approx 0$. The line in Figure~\ref{Fig_TrionExciton}(a) corresponds to a fit using $V_\mathrm{HOM} = M_\mathrm{s} - (1 + M_\mathrm{s} ) \gtwo$ from which we extract as a single parameter the single photon indistinguishability $M_\mathrm{s} = 0.920 \pm 0.003 $. 

For the trion based source, since the imperfect $\gtwo$ arises from re-excitation, the assumption of separable noise does not hold because the emission of the first and second photon are time correlated. However, the extra photon must be emitted during the laser pulse for re-excitation to occur~\cite{Fischer2017}, whereas the main single photon emission typically takes place after the laser pulse with the trion radiative decay time of approximately 170~ps. As a result the noise photon, while emitted by the QD, presents a very similar temporal profile to the laser photon and can be modelled as a temporally separated noise with $M_\mathrm{sn} = 0$. The validity of this analysis is verified by increasing the pulse duration to increase the probability of re-excitation. The $\gtwo$ and HOM visibility are measured for different pulse durations from 15~ps to 50~ps at the power that maximises the emitted count rate. The results shown in Figure~\ref{Fig_TrionExciton}(b) are modelled well using  $V_\mathrm{HOM} = M_\mathrm{s} - (1 +  M_\mathrm{s}) \gtwo $ with a single parameter $M_\mathrm{s} = 0.942 \pm 0.004$. 

To summarize, we find that, despite their different physical origins, the multi-photon component of both exciton and trion based QDSPSs can be treated as separable distinguishable noise. In the limit of low $\gtwo$, the single photon indistinguishability can thus be obtained using:

\begin{equation}
    M_\mathrm{s } = \frac{ V_\mathrm{HOM } + \gtwo}{1 - \gtwo}.\label{Eqn_CorrectionFactor}
\end{equation}

\noindent This correction factor can be applied to any QDSPS with a small $\gtwo$ and fast excitation pulse, in order to extract the intrinsic single photon indistinguishability, $M_\mathrm{s}$, given a measurement of $\gtwo$ and $V_\mathrm{HOM}$. The more general case where the HOM beam splitter has an intensity reflectivity $R$ and transmission $T$ is given in the Supplementary Material.

Interestingly, Equation~\ref{Eqn_CorrectionFactor} results in a higher single-photon indistinguishability than the one obtained using the identical noise model. It is therefore likely that many of the values of single photon indistinguishability that have been quoted in the literature~\cite{Senellart2017} for QD sources are in fact an underestimate of the true value. Whilst this does not circumvent the fact that it is the overall wavepacket indistinguishability that is crucial for quantum technologies, this deepened understanding of the Hong-Ou-Mandel experiment allows for a better diagnosis regarding imperfect single photon sources. The single-photon indistinguishability, $M_\mathrm{s}$, gives the upper bound to the indistinguishability that could be achieved with an ideal experimental set--up, with no laser leakage for example, and therefore it fundamentally quantifies how temporally coherent the source itself is. Finally, we note that our simple theoretical approach is only valid for separable noise, and further studies are needed to understand the effect of a non-separable noise on the interference.

In conclusion, we have theoretically and experimentally revisited the emblematic Hong-Ou-Mandel interference. This experiment is commonly implemented to test the indistinguishability of single particles including  single photons, single plasmons, single electrons or single atoms ~\cite{Martino2014,Bocquillon1054,Lopes2015}.   We believe that the new insight brought by our study will benefit these fundamental studies as well as the development of single photon sources, allowing a better diagnosis on the current limitations. 
\vspace{3mm}

\textit{Acknowledgements} This work was partially supported by the ERC PoC PhoW, the French Agence Nationale pour la Recherche (grant ANR QuDICE), the IAD - ANR support ASTRID program Projet ANR-18-ASTR-0024 LIGHT, the QuantERA ERA-NET Cofund in Quantum Technologies, project HIPHOP, the French RENATECH network, a public grant overseen by the French National Research Agency (ANR) as part of the "Investissements d’Avenir" programme (Labex NanoSaclay, reference: ANR-10-LABX-0035). J.C.L. and C.A. acknowledge support from Marie SkłodowskaCurie Individual Fellowships SMUPHOS and SQUAPH, respectively. H. O. Acknowledges support from the Paris Ile-de-France Région in the framework of DIM SIRTEQ. S.C.W. and C.S. acknowledge support from NSERC (the Natural Sciences and Engineering Research Council), AITF (Alberta Innovates Technology Futures), and the SPIE Education Scholarships program.

\bibliography{HOMg2}

\clearpage
\onecolumngrid 
	
	\appendix
	\setcounter{figure}{0} \renewcommand{\thefigure}{S.\arabic{figure}}
	\setcounter{equation}{0} 
	\renewcommand{\theequation}{S.\arabic{equation}}

\begin{center}
    \large{Supplementary Material}
\end{center}

\section{HOM visibility for unentangled states}

In this section, we derive the general relationship between the total mean wavepacket overlap of two interfering states, the single-photon purity, and the HOM visibility. Suppose we have a beam splitter with input modes $\hat{a}_1$ and $\hat{a}_2$ and output modes $\hat{a}_3$ and $\hat{a}_4$ monitored by single-photon detectors. The output modes can be described by the relation
\begin{equation}
\label{bsrelation}
    \begin{pmatrix}
    \hat{a}_3(t)\\
    \hat{a}_4(t)
    \end{pmatrix}
    =
    \begin{pmatrix}
    \cos\theta&-e^{-i\phi}\sin\theta\\
    e^{i\phi}\sin\theta&\cos\theta
    \end{pmatrix}
    \begin{pmatrix}
    \hat{a}_1(t)\\
    \hat{a}_2(t)
    \end{pmatrix},
\end{equation}
provided that the beam splitter interaction is constant for the relevant frequency range of the input fields. The coincident events of the detectors at the output are determined from the two-time intensity correlation between the output fields $G_{34}^{(2)}(t,\tau)=\braket{\hat{a}^\dagger_3(t)\hat{a}^\dagger_4(t+\tau)\hat{a}_4(t+\tau)\hat{a}_3(t)}$, where we restrict $\tau\geq0$ so that the detector monitoring mode $4$ clicks after the detector monitoring mode $3$. The integrated $G_{34}^{(2)}$ around zero delay normalized by the total intensity gives the probability $p_{34}$ of having a coincident count. The HOM interference visibility is then defined as $V_\text{HOM}=1-2p_{34}$. By following the methods of Ref. \cite{Imamoglu2004} to compute $p_{34}$ for two unentangled input states, while also keeping terms associated with multi-photon contributions, the HOM visibility is given by
\begin{equation}
    V_\text{HOM} = \frac{2RT\left[\left(1-g_1^{(2)}\right)\mu_1^2+2M_{12}\mu_1\mu_2+\left(1-g_2^{(2)}\right)\mu_2^2\right]}{(T\mu_1+R\mu_2)(T\mu_2+R\mu_1)}-1, 
\end{equation}
where $R=\sin^2\theta$ is the beam splitter reflectivity, $T=\cos^2\theta$ is the beam splitter transmittance,
\begin{equation}
    g_i^{(2)} = \frac{2\iint G_{ii}^{(2)}(t,\tau)d\tau dt}{\mu_i^2}=\frac{2\iint\braket{\hat{a}^\dagger_i(t)\hat{a}^\dagger_i(t+\tau)\hat{a}_i(t+\tau)\hat{a}_i(t)}d\tau dt}{\mu_i^2}
\end{equation}
is the integrated intensity correlation around zero delay for input $i$ normalized by the time-integrated mean photon number $\mu_i = \int\braket{\hat{a}^\dagger_i(t)\hat{a}_i(t)}dt$
and the mean-wavepacket overlap is
\begin{equation}
    M_{ij} = \frac{2\iint\text{Re}\left(\braket{\hat{a}^\dagger_i(t+\tau)\hat{a}_i(t)}^*\braket{\hat{a}^\dagger_j(t+\tau)\hat{a}_j(t)}\right)d\tau dt}{\mu_i\mu_j}.
\end{equation}

For a balanced interferometer where the input intensities of the final beam splitter are equal, we have that $\mu_1=\mu_2$ and so the relation for HOM visibility simplifies to
\begin{equation}
    V_\text{HOM} = 4RT\left(M_{12}+1-\overline{g}^{(2)}\right)-1,
\end{equation}
where $\overline{g}^{(2)}=(g_1^{(2)}+g_2^{(2)})/2$ is the average $g^{(2)}$ of the interfering states. For HOM interference between identical wavepackets, $M_{12} = M_{11}=M_{22}= M_\text{tot}$ is the total mean wavepacket overlap of the source and $\overline{g}^{(2)}=g^{(2)}_1=g^{(2)}_2=g^{(2)}$ quantifies the source single-photon purity. In the ideal case where $R=T=1/2$, the above relation reduces to the equation $V_\text{HOM}=M_\text{tot}-g^{(2)}$ given in the main text.

\section{HOM visibility in a separable noise model}

\begin{figure}[h]

	\includegraphics[width=0.3\linewidth]{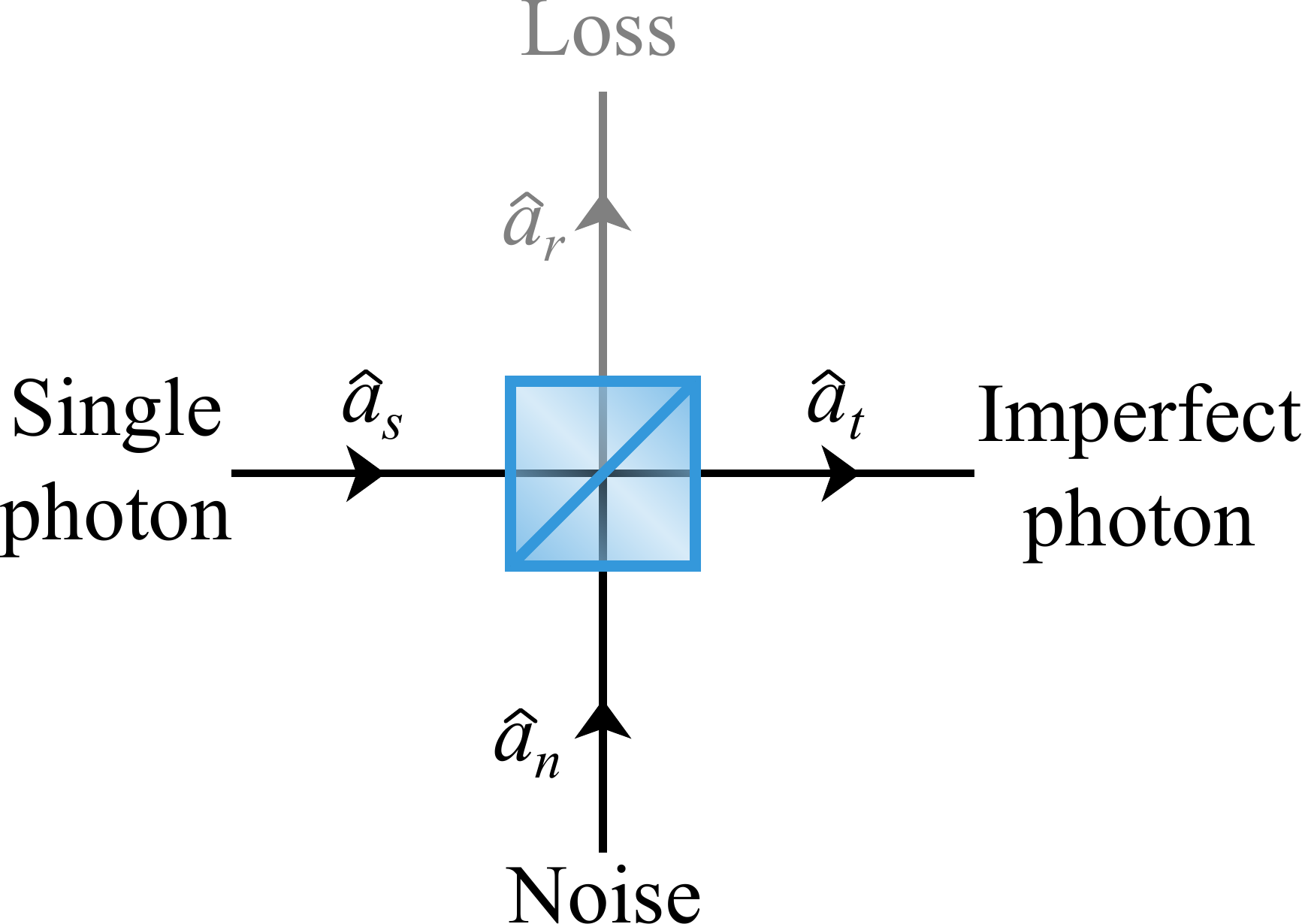}
	\caption{ 	{ An imperfect photon is modelled by adding separable noise to a perfect single photon at a beam splitter } \label{Fig_BS}}

\end{figure}

We now develop a theoretical model to describe the visibility of HOM interference for a specific type of imperfect source. This imperfect single photon source is modeled by adding noise to an ideal single photon using a beam splitter interaction, as shown in Figure~\ref{Fig_BS}. Here the noise is separable and exhibits no entanglement with the single photon. For simplicity, we also model the noise by another single photon, which is valid in the limit of a weak noise field or, equivalently, when $g^{(2)}$ is small.
	
The initial state of the signal ($s$) and noise ($n$) is given by $\hat{\rho} =\hat{\rho}_{s} \otimes \hat{\rho}_{n}  $ where $\hat{\rho}_i=p_{i,0}\ket{0}\bra{0}+p_{i,1}\hat{\rho}_{i,1}$ and
 \begin{equation}
    \hat{\rho}_{i,1} = \iint \xi_i(t,t') \hat{a} ^\dagger _i (t)\ket{0} \bra{0} \hat{a} _i (t') \, \mathrm{d}t \mathrm{d}t' \nonumber,
\end{equation}
where $\xi_i(t,t')$ is the normalized single-photon temporal density wavefunction for $i \in \{ s,n\} $.

The imperfect photon in the output transmission mode of the beam splitter $\hat{a}_t$ is obtained by tracing out the reflected loss mode $\hat{a}_r$ after applying the beam splitter relation, so that $\hat{\rho}_t = \mathrm{Tr}_r(\hat{\rho}_{s} \otimes \hat{\rho}_{n}) $, where 
\begin{equation}
   \begin{pmatrix} \hat{a}_t(t) \\ \hat{a}_r(t) \end{pmatrix} =  \begin{pmatrix} \cos\vartheta & -\sin\vartheta \\ \sin\vartheta & \cos \vartheta \end{pmatrix} \begin{pmatrix} \hat{a}_s(t) \\ \hat{a}_n(t) \end{pmatrix}.
\end{equation}

The total state of the imperfect single photon can then be written as $\hat{\rho}_t = p_{0}\ket{0}\bra{0} + p_{1}\hat{\rho}_{t,1} +p_{2}\hat{\rho}_{t,2} $, where $\hat{\rho}_{t,j}$ is the density matrix for the transmitted state containing $j$ photons. The total $g^{(2)}$ and $\mu$ can be computed directly from the photon number probabilities by $\mu = p_1 + 2p_2$ and $g^{(2)}=2p_2/\mu^2$. For the average photon number, we have $\mu = p_{s,1}\cos^2\vartheta+p_{n,1}\sin^2\vartheta$ and $g^{(2)}$ is given by
\begin{equation}
\label{separableg2}
\begin{aligned}
    \mu^2g^{(2)} &= 2p_{s,1}p_{n,1}(1+M_\mathrm{sn})\cos^2\vartheta\sin^2\vartheta,
\end{aligned}    
\end{equation}
where $M_\mathrm{sn} = \iint \mathrm{Re} ( \xi_s (t,t') \xi_n^*(t,t') ) \mathrm{d} t \mathrm{d} t'$ is the mean wavepacket overlap of the single photon and noise, and $\vartheta$ quantifies the amount of noise.

After applying a propagation phase $\phi_i$ to each of the photon density wavefunctions, the two-time amplitude correlation of $\hat{\rho}_t$ is given by
\begin{equation}
        \braket{\hat{a}^\dagger_t(t^\prime)\hat{a}_t(t)}=p_{s,1}\cos^2\vartheta\xi_s(t,t^\prime)e^{i\phi_s(t-t^\prime)}+p_{n,1}\sin^2\vartheta\xi_n(t,t^\prime)e^{i\phi_n(t-t^\prime)}
\end{equation}
and so the total mean wavepacket overlap is given by
\begin{equation}
\label{separableoverlap}
\begin{aligned}
\mu^2M_\text{tot} &= \iint\left|\braket{\hat{a}^\dagger_t(t^\prime)\hat{a}_t(t)}\right|^2dt^\prime dt\\
&= p_{s,1}^2M_\mathrm{s}\cos^4\vartheta + p_{n,1}^2 M_n\sin^4\vartheta + 2p_{s,1}p_{n,1}M_\mathrm{sn}^\prime\cos^2\vartheta\sin^2\vartheta,
\end{aligned}
\end{equation}
where $M_{s} = \iint\left|\xi_s(t,t^\prime)\right|
    ^2 dtdt^\prime = \text{Tr}\left[\hat{\rho}_{s,1}^2\right]$
quantifies the intrinsic single-photon indistinguishability, or single-photon trace purity \cite{Fischer2018,Trivedi2020}, of the source. Here we have that
\begin{equation}
    M_\mathrm{sn}^\prime = \iint \text{Re}\left(\xi_s(t,t^\prime)\xi_n^*(t,t^\prime)e^{i(\phi_s-\phi_n)(t-t^\prime)}\right) dtdt^\prime
\end{equation}
is not necessarily the same as $M_\mathrm{sn}$ due to the potential relative propagation phase $\phi_s-\phi_n$.

We can reparametrize the expressions for $\mu$, $g^{(2)}$, and $M_\text{tot}$ by defining $\eta$ so that $\cos^2\!\eta=(p_{s,1}\cos^2\vartheta)/\mu$ and $\sin^2\!\eta=(p_{n,1}\sin^2\vartheta)/\mu$. The fact that this reparametrization exists stems from the independence of $M_\text{tot}$ and $g^{(2)}$ from photon loss. It also implies that the fundamental quantity affecting the photon statistics of this imperfect single photon model is $\eta$, which depends on both the beam splitter angle $\vartheta$ and the relative input intensities through $p_{s,1}$ and $p_{n,1}$. Using the relation for $V_\text{HOM}$ from the previous section, equations (\ref{separableg2}), and (\ref{separableoverlap}), the visibility and $g^{(2)}$ in terms of the noise parameter $\eta$ are
\begin{equation}
\begin{aligned}
    V_\text{HOM}(\eta) &= 4RT\left(1+M_\mathrm{s}\cos^4\eta+M_n\sin^4\eta -2(1+M_\mathrm{sn}-M_\mathrm{sn}^\prime)\cos^2\eta\sin^2\eta\right)-1\\
    g^{(2)}(\eta) &= 2(1+M_\mathrm{sn})\cos^2\eta\sin^2\eta.
\end{aligned}
\end{equation}
The value of $\eta$ can be modified by changing the intensity of the noise $p_{n,1}$ as was done for the emulated distinguishable noise source in the main text. It can also be modified by changing the relative proportions of $p_{n,1}$ and $p_{s,1}$ using an unbalanced Mach-Zehnder interferometer, as was done for the emulated identical noise source in the main text.

In our study, we are interested in the slope and intercept of the parametric curve formed by $\{g^{(2)}(\eta),V_\text{HOM}(\eta)\}$. The solution for the intercept is clear since $g^{(2)}(\eta)=0$ implies $\eta=0$ and $V_\text{HOM}(0) = 4RT(1+M_\mathrm{s})-1$. To solve for the slope at small $\eta$, we have
\begin{equation}
\begin{aligned}
    \lim_{\eta\rightarrow 0}\frac{dV_\text{HOM}(\eta)}{dg^{(2)}(\eta)}
    &= -4RT\left(\frac{1+M_\mathrm{s} + (M_\mathrm{sn}-M_\mathrm{sn}^\prime)}{1+M_\mathrm{sn}}\right).
\end{aligned}
\end{equation}

For the cases we are interested in, either distinguishable noise or if $\xi_s=\xi_n$, we have $M_\mathrm{sn}-M_\mathrm{sn}^\prime\simeq 0$. This case would also be true if $M_\mathrm{sn}$ and $M_\mathrm{sn}^\prime$ were phase averaged. Under these conditions, the HOM visibility for small $g^{(2)}$ is given by
\begin{equation}
\label{vhomRT}
    V_\text{HOM} = 4RT\left(1+M_\mathrm{s}-\left(\frac{1+M_\mathrm{s}}{1+M_\mathrm{sn}}\right)g^{(2)}\right)-1.
\end{equation}

In the case where the noise is distinguishable so that $M_\mathrm{sn}=0$, the single-photon indistinguishability $M_\mathrm{s}$ can be determined by rearranging equation (\ref{vhomRT}): 

\begin{equation}
    M_\mathrm{s } = \frac{ V_\mathrm{HOM}+4RT\left(1+\gtwo\right)-1}{4RT(1 - \gtwo)}
\end{equation}

For $R=T=1/2$, we recover equations (1) and (2) presented in the main text.

\section{Measuring $\gtwo$ and HOM interference} 
	
\begin{figure}[h]

	\includegraphics[width=0.53\linewidth]{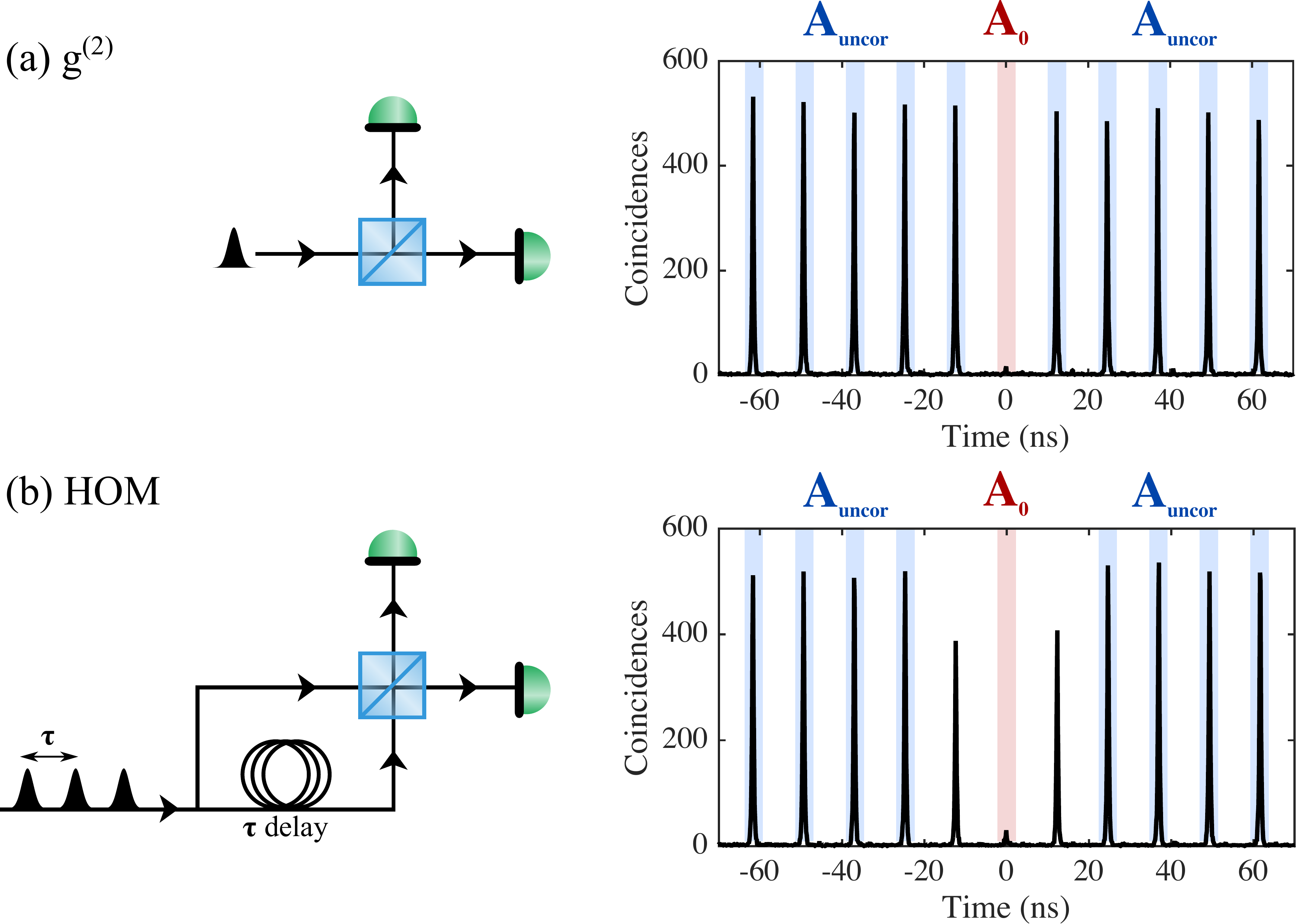}
	\caption{ 	{ The experimental setup (left) and raw data (right) for measuring (a) $\gtwo$ and (b) HOM visibility.  } \label{Fig_g2HOM_setup_data}}

\end{figure}

To measure the single photon purity we perform a Hanbury Brown-Twiss experiment and measure the coincidences between the two outputs of a 50:50 beam splitter. The experimental set-up, and a typical coincidence histogram are shown in Figure~\ref{Fig_g2HOM_setup_data}(a). The second-order autocorrelation is given by $\gtwo = A_0 / A_\mathrm{uncor}$, where $A_0$ is the area of the coincidence peak at zero time delay and $A_\mathrm{uncor}$ is the average area of the uncorrelated peaks. 

We perform a Hong-Ou-Mandel interference experiment by splitting the train of single photons at a beam splitter and delaying one arm by the pulse separation time, $\tau$. Two subsequently emitted photons then interfere at a 50:50 beam splitter, as shown in Figure~\ref{Fig_g2HOM_setup_data}(b). The visibility of HOM interference is given by $V_\mathrm{HOM} = 1 - 2 A_0 / A_\mathrm{uncor}$.

\end{document}